\documentclass[aps,twocolumn,floatfix,showpacs, nofootinbib]{revtex4}
\usepackage{graphicx,bm,hhline, amsmath }

\newcommand{\cf}{{\cal F}}
\newcommand{\br}{{\bm r}}

\newcommand{\brp}{{\br}^\prime}

\newcommand{\ul}{\underline}

\begin{document}
\title{An integral equation model for warm and hot dense mixtures}
\author{C. E. Starrett, D. Saumon and J. Daligault}
\affiliation{Los Alamos National Laboratory, P.O. Box 1663, Los Alamos, NM 87545, U.S.A.}
\author{S. Hamel}
\affiliation{Lawrence Livermore National Laboratory, Livermore, CA 94550, U.S.A}

\date{\today}
\begin{abstract}
In Starrett and Saumon [Phys. Rev. E {\bf 87}, 013104 (2013)] a model for the calculation
of electronic and ionic structures of warm and hot dense  matter was described and validated.  
In that model the electronic structure of one `atom' in a plasma is determined using a density
functional theory based average-atom (AA) model, and the ionic structure is determined by 
coupling the AA model to integral equations governing the fluid structure.
That model was for plasmas with one nuclear species only.  Here we extend it to treat plasmas
with many nuclear species, i.e. mixtures, and apply it to a carbon-hydrogen mixture relevant to
inertial confinement fusion experiments.  Comparison of the predicted electronic and ionic structures with 
orbital-free and Kohn-Sham molecular dynamics simulations reveals excellent agreement wherever chemical 
bonding is not significant.
\end{abstract}
\maketitle

\section{Introduction}
Warm and hot dense matter refers to plasmas roughly as dense as solids up to thousands of times solid
density.  Temperatures range from approximately an eV up to several thousand eV.  In nature such plasmas
are found in a variety of astrophysical objects, including the cores of giant planets and in the envelopes of white
dwarfs \cite{HEDLP_report, iau147}.  Such plasmas are also of interest to the inertial confinement 
fusion community where these conditions are reached in implosion experiments \cite{hammel}.
Often the plasmas in question are mixtures.  For example, CH$_{1.36}$\footnote{The notation means a carbon-hydrogen mixture
in the ratio 0.424:0.576.} is of interest as an ablator material in inertial
confinement fusion experiments at the National Ignition Facility (NIF) \cite{hamel-ch} and carbon/helium mixtures
are of interest to the white dwarf community \cite{HEDLP_report, iau147}.

One of the challenges of modeling warm and hot dense matter is to accurately calculate the
electronic and ionic structures in a single model.  Benchmark calculations have been made 
using density functional theory molecular dynamics (DFT-MD).  This method is thought to be accurate
and gives an essentially complete description of the plasma.  Both Kohn-Sham (KS) \cite{mazevet1,desjarlais2}
and orbital-free (OF) \cite{zerah}
versions of DFT-MD exist.  The primary limitation of these methods is their high computational
cost.  This is particularly acute for KS calculations, where a poor scaling of the computational cost with temperature
limits it to low temperatures.  The OF method does not suffer from this poor scaling but is still very
expensive, typically limiting the number of particles in the MD simulation to a few 100's \cite{danel06, danel08, danel12, arnault13}.
DFT-MD calculations are especially challenging for mixtures, where asymmetries in masses and number
fractions increase the computational demands.

In \cite{starrett13, starrett14} an alternative method for calculating the electronic and ionic 
structures of warm and hot dense matter, in the form of pair distribution functions, was presented.  Excellent agreement
was found with corresponding DFT-MD simulations over a wide range of densities and temperatures.  
The principal advantage of this model is that it is much
less computationally expensive than the corresponding DFT-MD simulations.  The model uses
a DFT based, average-atom (AA) approach to calculate the properties of one `atom' in the plasma and couples
this to the quantum Ornstein-Zernike (QOZ) equations for the ionic structure.  The QOZ's are integral
equations that can be rapidly solved, giving the ion-ion and ion-electron pair distribution functions.
The AA model can be solved using either KS or OF functionals.  
A key assumption of the model is that the electronic density of the plasma can be written as a superposition 
of ``pseudoatom'' electron densities.  The concept of the pseudoatom \cite{perrot1, dagens72, ziman67}
is that of a charge neutral, atom-like entity with a nuclear charge at its origin, surrounded by a local, spherically 
symmetric electron cloud.  This electron cloud comprises the electrons that are bound to the nucleus (together with 
the nucleus these form the ion), as well as screening electrons.  
The model of \cite{starrett13, starrett14} is, however, limited to homo-nuclear plasmas.  Here we extend this model
to hetero-nuclear plasmas, i.e. mixtures.  

We first extend the QOZ equations to 
mixtures of quantal electrons and N types of classical ions.  The result turns out to be a straightforward 
generalization of the corresponding N-component classical OZ equations, though the derivation
is not trivial.  Secondly, we show how the average-atom model developed in \cite{starrett13, starrett14}
can be coupled to the QOZ's for mixtures.  
Lastly, we present an application of the model to CH$_{1.36}$ and 
compare the resulting electronic and ionic structures to OFMD and QMD\footnote{Quantum Molecular dynamics.  This is 
DFT-MD with the orbital based Kohn-Sham method.} simulations.  

The resulting complete model allows rapid calculation of the electronic and ionic structures 
of dense plasma mixtures (relative to DFT-MD) with no adjustable parameters.  
The extension to mixtures does not require any new physical approximations.  Inputs
to the model are the ion data (masses and nuclear charges),  the number fractions of the ion species, as 
well as the plasma mass density and temperature.  
Lastly, we note that the model is all-electron, meaning that, unlike DFT-MD simulations, no pseudopotential is used
when solving for the electronic structure.

The structure of this paper is as follows.  In section \ref{sec2} we develop the QOZ equations for arbitrary\footnote{Arbitrary meaning
any number fraction and mass or charge ratio.}
multi-component fluids.  We start by deriving the QOZ equations for a binary mixture (two classical
ion species and quantal electrons).  The extension to the N-component mixture is then obvious, and
analogous to the corresponding classical OZ equations \cite{hansen1}.  We then show how these equations can
be written in terms of electronic screening densities of pseudoatoms, in analogy with the homo-nuclear
case.  The next step is to demonstrate how the QOZ equations can be mapped onto an effective N-component
system, where the ions interact through short ranged, electron screened, effective potentials.  Finally,
we show how to calculate the screening densities for mixtures using the average-atom model developed
in \cite{starrett13, starrett14}.  
In section \ref{sec3} we compare predictions of the electronic and ionic structures, in the form of pair distribution functions,
to both quantum and orbital-free molecular dynamics simulations (QMD and OFMD).  
Finally, in section \ref{sec4} we present our conclusions.  Unless otherwise stated, atomic units, in which 
$\hbar = m_e = k_{B} = e = 1$, where the symbols have their usual meaning, are used throughout.

\section{The quantum Ornstein-Zernike equations for mixtures}
\label{sec2}
We consider a mixture of $N$ types of classical species with a neutralizing, responding, electron gas.
The electrons and ions are in thermal equilibrium with temperature $T=1/\beta$.
The number fraction for ions of type $i$ is $x_i$, such that
\begin{equation}
\sum\limits_{i=1}^N x_i = 1.
\end{equation}
If the charge of ion $i$ is $\bar{Z}_i$, the average ionization of the plasma is
\begin{equation}
\bar{Z} = \frac{\bar{n}_e^0}{n_I^0} = \sum\limits_{i=1}^N \bar{Z}_i x_i
\end{equation}
where $\bar{n}_e^0$ is the average ionized electron particle density and $n_I^0$ is the 
average ion particle density.  We also define the particle density for species $i$
as
\begin{equation}
n_i^0 = x_i\, n_I^0
\end{equation}

\subsection{The Quantum Ornstein-Zernike matrix}
Chihara \cite{chihara84a} derived the quantum Ornstein-Zernike (QOZ) equations
for a mixture of one classical ion species and quantum mechanical electrons.  Here 
we extend this derivation to a multi-component mixture of $N$-classical ions and quantum 
mechanical electrons. We start from the exact matrix equation \cite{chihara84a} in
$k$-space (see appendix \ref{ft})
\begin{equation}
\ul{\chi} = \left[ \frac{\ul{C}}{\beta} + \left(\ul{\chi}^{(0)}\right)^{-1} \right]^{-1}
\label{matrix1}
\end{equation}
where the underline indicates a matrix.  This formula relates the linear response functions $\ul{\chi}$ 
for an interacting system to those of a non-interacting system $\ul{\chi}^0$ and the direct correlation
functions $\ul{C}$.   For homogeneous system these are defined in real space as
\begin{equation}
\chi_{ij}(| \br - \br^\prime| )
= - \frac{ \delta^2 \Omega }{\delta \Phi_i(\br) \delta \Phi_j(\brp) }
\end{equation}
and
\begin{equation}
  \frac{-1}{\beta} C_{ij}(\mid \br-\brp \mid) \equiv
  \left.\frac{\delta^2\cf^{ex}}{\delta n_i(\br) \delta n_j(\brp)}\right|_{V_i=0}
\end{equation}
where 
\begin{equation}
\Phi_i(\br) = \mu_i - V_i(\br)
\end{equation}
is the intrinsic chemical potential for species $i$ with particle density $n_i(\br)$, chemical potential $\mu_i$
and external potential $V_i(\br)$.  $\Omega$ is the grand potential
\begin{equation}
\Omega = \cf - \sum\limits_{i}^{N+1} \int d\br  \Phi_i(\br)  n_i(\br) 
\end{equation}
and $\cf$ is the intrinsic free energy and 
\begin{equation}
\cf =  \cf^{id} + \cf^{ex}.
\end{equation}
$\cf^{id}$ is the non-interacting intrinsic free energy and $\cf^{ex}$ is the intrinsic
free energy due to interactions \cite{hansen1}.  The notation $V_i=0$ indicates that the 
functional derivative is evaluated with the external potential set to zero.   
Finally, the matrices have size $(N+1)\times(N+1)$ for a system of electrons and $N$ ion species
and are symmetric.  

To solve the matrix equation (\ref{matrix1}) we specialize to a mixture of quantal electrons (index $e$) and 2 classical ion species (indices 1 and 2).
This greatly simplifies the problem and the solution to the general $(N+1)$ problem can be inferred from
the result.  We can use the fluctuation-dissipation theorem to relate the response functions $\chi$ to the
corresponding structure factors $S_{ij}$ \cite{hansen1} provided one of the particles ($i$ or $j$) is classical,
\begin{eqnarray}
\chi_{ij}(k) & = &-\beta \sqrt{n_i^0 n_j^0} S_{ij}(k) \nonumber\\
             & = &-\beta \sqrt{n_i^0 n_j^0} \left[ \delta_{ij} + \sqrt{n_i^0 n_j^0} h_{ij}(k) \right]
\end{eqnarray}
where $\delta_{ij}$ is the Kronecker delta and $h_{ij}(k)$ is a pair correlation function.
The (3$\times$3) matrix in square brackets in equation (\ref{matrix1}) is now inverted, giving
the QOZ equations for a binary mixture of classical ions and quantum electrons:
\begin{eqnarray}
\!\!h_{ij}(k)\! &\!\! =\!\! &\!\! \left(  \frac{\chi_{jj}^{0}(k)}{-\beta n_j^0} \right) 
\left[ C_{ij}(k) + \sum\limits_{\lambda=1}^{3} n_\lambda^0 h_{i\lambda}(k) C_{\lambda j}(k)\right]
\label{qoz_binary}
\end{eqnarray}
where the index $\lambda = 3$ is for the electrons
and the convention is that if either the $i$ or $j$ labels refer to an electron, then it is placed in the
$j$ position in equation (\ref{qoz_binary}) (recall the symmetry $h_{i j} = h_{j i}$).  

By simple extension we can now write down the equations for the $(N+1)$ component plasma ($N$ classical particles
and quantum electrons)
\begin{eqnarray}
\!\!h_{ij}(k)\! &\!\! =\!\! &\!\! \left(  \frac{\chi_{jj}^{0}(k)}{-\beta n_j^0} \right) 
\left[ C_{ij}(k) + \sum\limits_{\lambda=1}^{N+1} n_\lambda^0 h_{i\lambda}(k) C_{\lambda j}(k)\right]
\label{qoz_general}
\end{eqnarray}
where $\lambda=N+1$ is the index for the electrons.  Equation (\ref{qoz_general}) can be compared 
to the familiar expression for a mixture of $(N+1)$ classical particles \cite{hansen1} (see also equation (\ref{coz}).
The quantum nature of the electrons is embodied in the pre-factor $\chi_{ee}^{0}(k)/(-\beta \bar{n}_e^0)$.
We note that $\chi_{ee}^0$ is the well-known finite temperature, non-interacting response function \cite{chabrier}.  At zero
temperature it is the Lindhard function \cite{ichimaru1}.  For classical particles $\chi_{jj}^0(k)= -\beta n_j^0$ and the
prefactor equals unity.

\section{Interpretation of the QOZ equations as a system of screened ions}
\label{qoz_scr}
The induced electronic screening density $n_{i,e}^{scr}(r)$ due to a weak external potential
$-C_{ie}/\beta$ is given by linear response theory \cite{ashcroft78} as
\begin{eqnarray}
n_{i,e}^{scr}(k) = -\frac{C_{i e}(k)}{\beta} \chi_{ee}^\prime(k)
\label{scr_den_def}
\end{eqnarray}
where
\begin{equation}
\chi_{ee}^\prime(k) \equiv
\frac{\chi_{ee}^{0}(k) }
{ 1 + \chi_{ee}^{0}(k) C_{ee}(k) / \beta}.
\label{chip}
\end{equation}
From equation (\ref{qoz_general}) we therefore have
\begin{flalign}
\Delta n_{i,e}(k)  \equiv
\bar{n}_e^0 h_{ie}(k) 
 = & n_{i,e}^{scr}(k) + \sum\limits_{\lambda=1}^{N} n_\lambda^0 h_{i\lambda}(k) n_{\lambda, e}^{scr}(k)
\label{dnie}
\end{flalign}
where
\begin{eqnarray}
\Delta n_{i,e}(r) & = & n_{i,e}(r) - \bar{n}_e^0
\end{eqnarray}
and $n_{i,e}(r)$ is the spherically averaged electron density around an ion species $i$.  The formula (\ref{dnie})
says that for homogeneous plasmas and weak external potentials $C_{ie}(k)$, the electron density of the plasma is exactly
written as the sum of spherically symmetric screening densities $n_{i,e}^{scr}(r)$.
The QOZ equations can therefore be interpreted as relations for the structure of a fluid of classical ions whose interactions
are screened by responding electrons with densities $n_{i,e}^{scr}(r)$.

\section{Reduction to an effective $N$-component system of classical particles}
As in the homo-nuclear case, to solve the QOZ equations (\ref{qoz_general}) for the ion-electron
and ion-ion pair correlation functions $h_{ie}$ and $h_{ii}$ we make use
of the interpretation given section \ref{qoz_scr} to map the QOZ equations
onto their purely classical counterparts:
\begin{eqnarray}
h_{IJ}(k) & = & C_{IJ}(k) + \sum\limits_{\lambda=1}^{N} n_\lambda^0 h_{I\lambda}(k) C_{\lambda J}(k)
\label{coz}
\end{eqnarray}
This procedure has be extensively described for the homo-nuclear case \cite{anta, starrett13};
here we give only the salient details.  

We assume that there exists an effective $N$-component system of classical particles, interacting
through short ranged pair potentials $V_{IJ}(r)$ such that the ion-ion pair correlation functions $h_{IJ}$
are identical to those of a corresponding\footnote{Corresponding meaning that the classical particles have the same
charge, mass and number densities.} ($N$+1)-component system of classical particles and quantal 
electrons.  The ion-ion closure relation for the effective $N$-component system is
\begin{flalign}
h_{IJ}(r) + 1 & =  \exp\left(  -\beta V_{IJ}(r) + h_{IJ}(r) - C_{IJ}(r) + B_{IJ}(r) \right) \label{c_hij}
\end{flalign}
where $I,J=1,\ldots,N$ and $B_{IJ}$ are bridge functions \cite{hansen1}.  Similarly, for the ($N$+1) component system 
we have
\begin{flalign}
h_{ij}(r) + 1 & =  \exp\left(  -\beta \frac{\bar{Z}_i\, \bar{Z}_j}{r} + h_{ij}(r) - C_{ij}(r) + B_{ij}(r) \right) \label{c_hlcij}
\end{flalign}
where $i,j=1,\ldots,N$ and $\bar{Z}_i$ is the charge of ion $i$.  By assuming that $h_{ij}=h_{IJ}$ and that corresponding bridge functions 
are equal (i.e. $B_{ij}=B_{IJ}$), one uses the QOZ and OZ equations (\ref{qoz_general}) and (\ref{coz}) with equations 
(\ref{c_hij}) and (\ref{c_hlcij}) to relate the pair potentials to the screening densities:  
\begin{eqnarray}
V_{I J}(k) & = & \frac{ 4 \pi \bar{Z}_i \bar{Z}_j }{k^2} - \frac{C_{i e}(k)}{\beta} n_{j,e}^{scr}(k).
\label{coz_pot}
\end{eqnarray}
Recall that the ion-electron direct correlation functions $C_{i e}(k)$ are also related to the screening densities via 
equation (\ref{scr_den_def}).

The problem of solving the QOZ equations (\ref{qoz_general}) is now reduced to solving their purely classical counterparts 
(equations (\ref{coz})) for given pair potentials $V_{IJ}$, using the ion-ion closure relations (\ref{c_hij}).  An algorithm for
solving these equations is described in appendix \ref{noz_num}.  To determine the potentials (\ref{coz_pot}) we need both
the screening densities $n_{i,e}^{scr}(k)$ for each ion and the electron-electron direct correlation function
$C_{ee}(k)$ (equation (\ref{chip})).  For the latter we use the jellium approximation which was successful for the 
homo-nuclear case \cite{starrett14}.  To determine the screening densities we also use the same approximation as for
the homo-nuclear case, i.e. we determine them using an average-atom model.

\begin{figure*}[!]
\begin{center}
\includegraphics[scale=0.9]{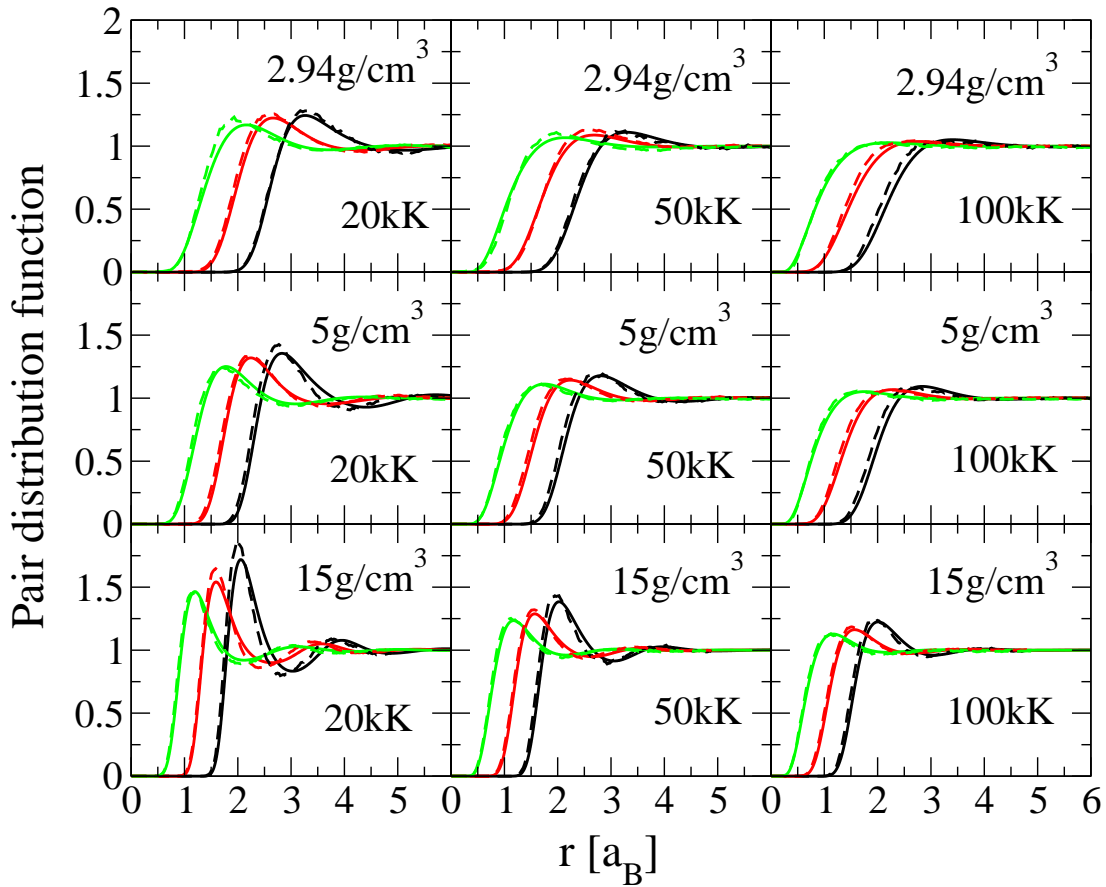}
\end{center}
  \caption{Ion-ion pair distribution functions for CH$_{1.36}$.  IS-TF (solid lines) compared to OFMD simulation
results (dashed lines) in the TF approximation.  C-C in black (right-most lines), C-H in red (middle lines) and H-H in
green (left-most lines).}
  \label{fig_ch_tf}
\end{figure*}
\begin{figure}[!]
\begin{center}
\includegraphics[scale=0.7]{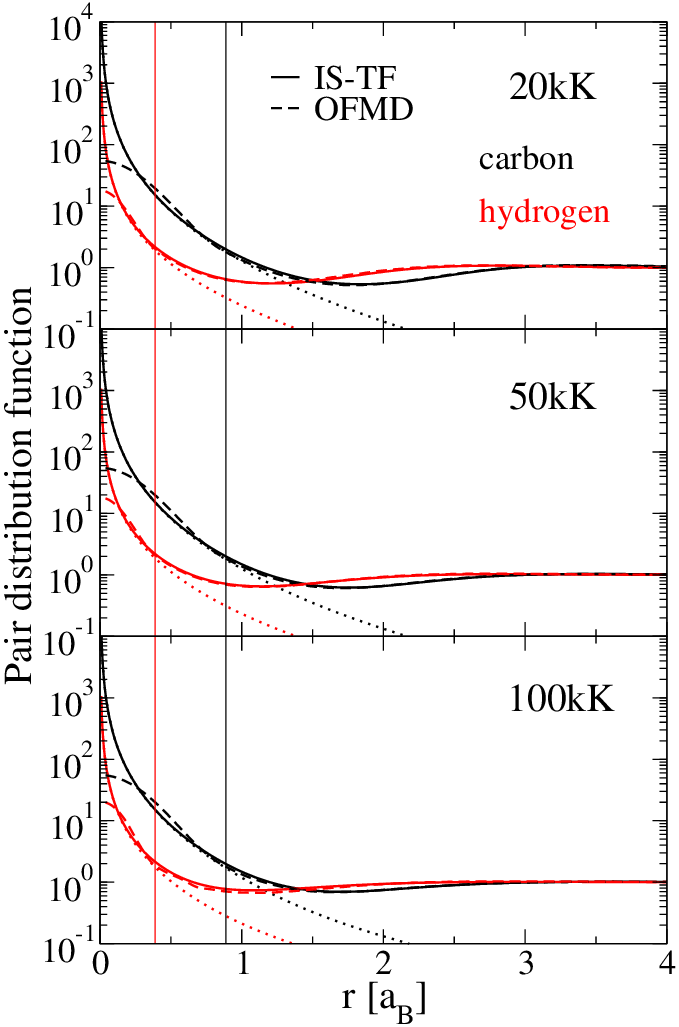}
\end{center}
  \caption{IS-TF ion-electron pair distribution functions for CH$_{1.36}$ at 2.94g/cm$^3$, compared to OFMD simulation
results in the TF approximation.  Carbon in black (right-most lines), hydrogen in red (left-most lines).  Solid lines are 
IS-TF, dashed lines are OFMD.  Also shown (dotted lines) is the contribution to $g_{ie}(r)$ from $n_e^{PA}(r)$ alone. 
The thin vertical lines indicate the cutoff radius $r_c$ used to generate the pseudopotentials for the OFMD simulations.}
  \label{fig_ch_tf_gie}
\end{figure}
%
%

\begin{figure*}[!]
\begin{center}
\includegraphics[scale=0.9]{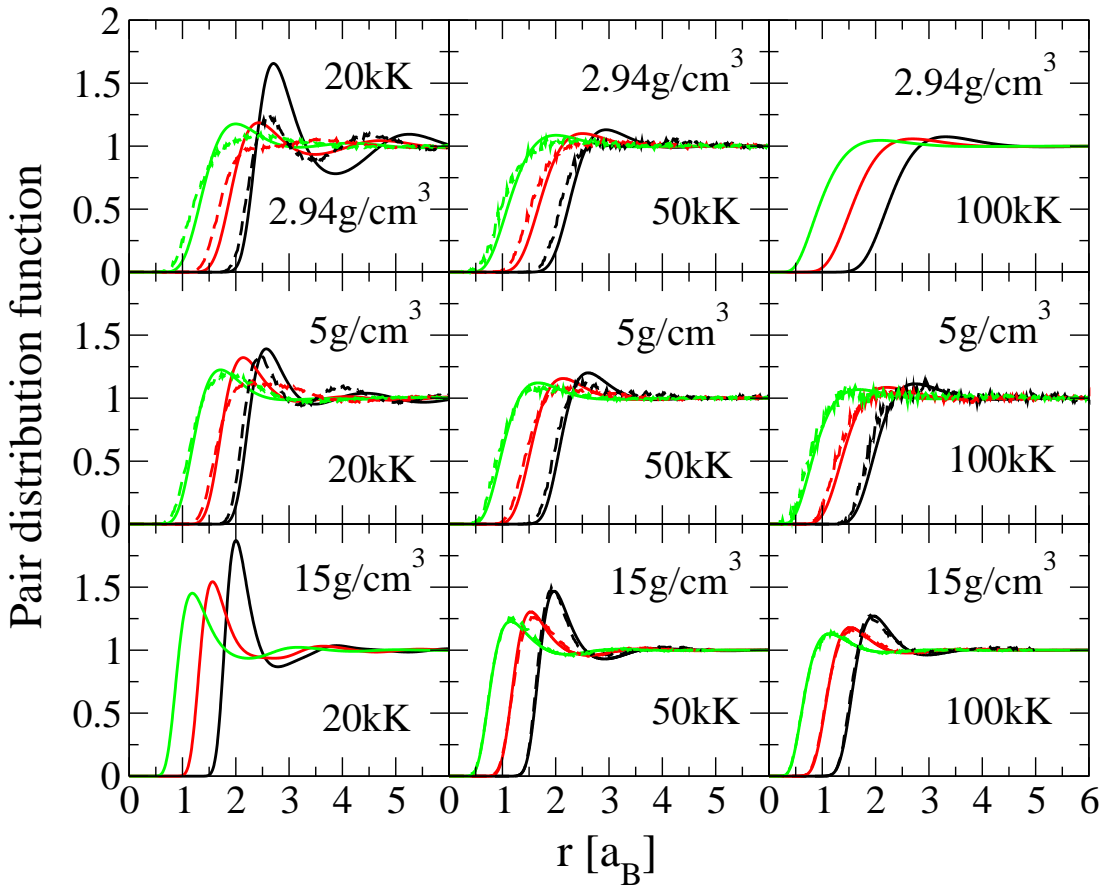}
\end{center}
  \caption{Ion-ion pair distribution functions for CH$_{1.36}$.  IS-QM (solid lines) compared to QMD simulations (dashed lines).  
C-C in black (right-most lines), C-H in red (middle lines) and H-H in
green (left-most lines).  For panels with only IS-QM curves, corresponding QMD results were not available.}
  \label{fig_ch2}
\end{figure*}
\begin{figure*}[!]
\begin{center}
\includegraphics[scale=0.9]{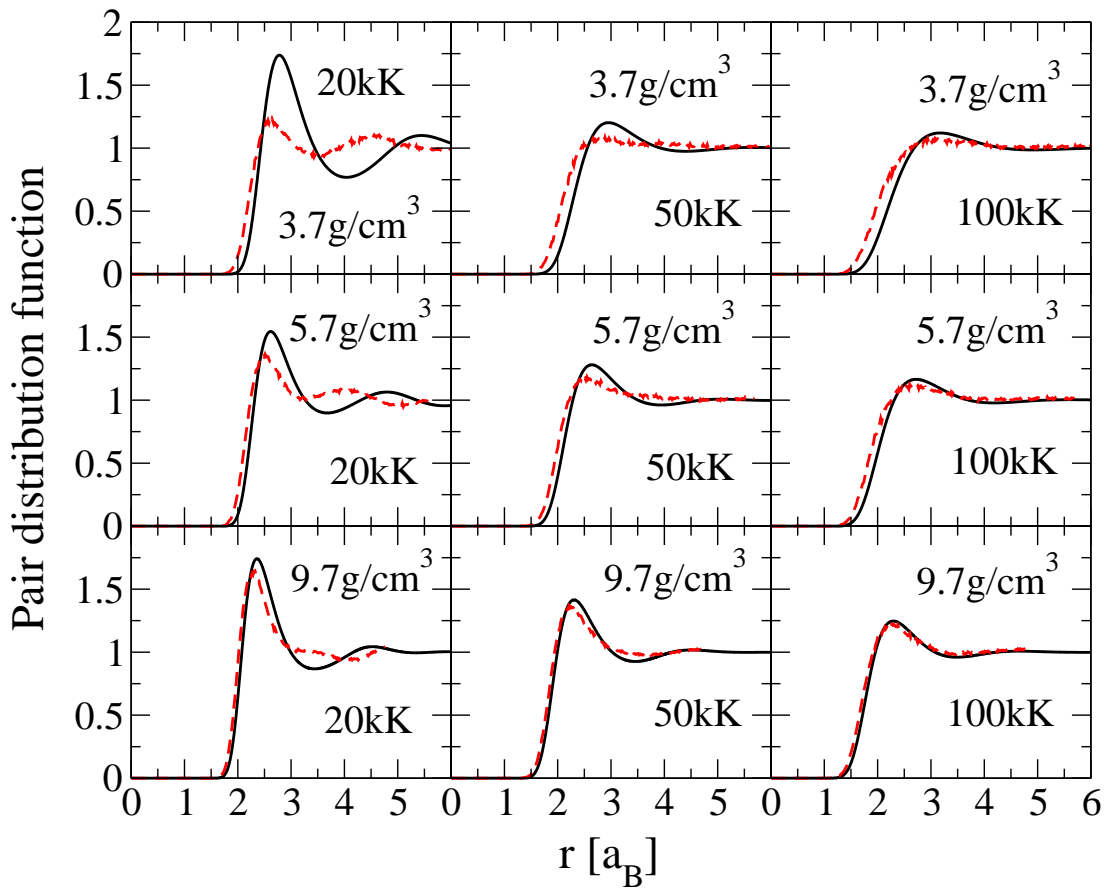}
\end{center}
  \caption{Ion-ion pair distribution functions for pure C.  IS-QM (solid lines) compared
to QMD simulations (dashed lines).}
  \label{fig_purec}
\end{figure*}

\section{The average-atom model for the screening density}
\label{sec_aa}
For the homo-nuclear case the method for calculating the screening density is described in detail 
in \cite{starrett13, starrett14}.   It is summarized here to provide a basis for its extension to mixtures.
We assume that the electron density of the plasma is given by a superposition of charge neutral `pseudoatom'
electron densities.  Conceptually, the pseudoatom electron density $n_e^{PA}(r)$ is a 
local, spherically symmetric electron cloud that contains both the bound states, with electron density $n_e^{ion}(r)$,
and the screening electrons ($n_e^{scr}(r)$) that contribute to the valence electrons.
Clearly, the superposition approximation will be excellent for deeply bound states, and equation (\ref{dnie})
demonstrates that it is also appropriate for the valence states provided these states can be
reasonably well represent by linear response theory.  It becomes inaccurate in 
thermodynamic regimes where a significant fraction of the atoms in the plasma form molecules.

To calculate $n_e^{PA}(r)$ we consider a system with a nucleus at the origin, surrounded by a spherically averaged
ion distribution (the `full' system).  The electron density $n_e^{full}(r)$ for this system is found by minimizing the
free energy for a given kinetic energy functional (eg. Kohn-Sham or Thomas-Fermi) and a given
exchange-correlation functional.  We then consider a system with the same spherically averaged
ion distribution but no central nucleus (the `external' system).  The electron density $n_e^{ext}(r)$
for this external system is found in the same way as for the full system.  The difference in these electron densities defines the
pseudoatom electron density 
\begin{equation}
n_{e}^{PA}(r) = n_{e}^{full}(r) - n_{e}^{ext}(r).
\end{equation}
To calculate the screening density from this we define a bound state electron density
$n_e^{ion}(r)$.  The screening density is then
\begin{equation}
n_{e}^{scr}(r) = n_{e}^{PA}(r) - n_{e}^{ion}(r).
\end{equation}

The above procedure requires a knowledge of the spherically averaged ion distribution.  This distribution
is given by the ion-ion pair distribution function $g(r)$($=h(r)+1$).  In the ion-sphere (IS) version
of the homo-nuclear model \cite{starrett14}, we solve for the screening density, as we have described
above, with $g(r)$ set to be a step function at the ion-sphere radius $R$
\begin{equation}
g(r) = \Theta(r-R)
\label{isgr}
\end{equation}
where 
\begin{equation}
\frac{4\pi R^3}{3} = V^{ion} = \frac{M}{\rho}
\label{ris}
\end{equation}
and $V^{ion}$ is the volume per ion, which is determined from input of the mass density $\rho$ and atomic mass $M$.
In \cite{starrett14} it was shown that the $g(r)$ from this IS model is very close to that of a self-consistent version
of the model, where the calculated $g(r)$ is fed back as input to the AA model, and the cycle repeated until converged.
This result was explained in \cite{starrett14} by noting 
that in the linear response regime $n_e^{scr}(r)$ is fully independent of $g(r)$.

The same technique for determining the screening densities can be applied here for mixtures, where now an 
AA model is solved for each component.  Because of the demonstrated insensitivity of $n_e^{scr}(r)$ to $g(r)$
we use the ion-sphere (IS) AA model \cite{starrett14}, both in the QM and TF versions of the model 
(referred to as IS-QM and IS-TF).  For species $i$ the ion-sphere radius $R_i$ is related the volume per
ion $V_i^{ion}$ but, unlike the homo-nuclear case, the volumes $V_i^{ion}$ are not uniquely determined from input.  
The volumes do however, satisfy 
\begin{eqnarray}
\sum\limits_{i=1}^N  x_i  V_i^{ion} & = & \frac{\sum\limits_{i=1}^{N} x_i\,M_i }{\rho}.
\label{vol_con}
\end{eqnarray}
where $M_i$ is the atomic mass of species $i$.
Further, the chemical potentials $\mu_e^i$ of the electrons for each average atom ($i=1,\ldots,N$) must be equal
\begin{equation}
\mu_e^i = \mu_e
\label{mu_con}
\end{equation}
The solution of the AA provides a relation between $\mu_e^i$ and $V_i^{ion}$.  Thus, the set of $V_i^{ion}$
are adjusted to satisfy equations (\ref{vol_con}) and (\ref{mu_con}), with $\mu_e$ a priori unknown.  The
solution can be sped up considerable by pre-tabulating $\mu_e^i$ as a function of $V_i^{ion}$ for the
species of interest.
This process yields a single electron chemical potential, bound and continuum wave functions and the nucleus-electron 
interaction potentials as well as the ion and screening densities for each ion species.

\section{Pair distribution functions}
The above model allows us to determine the ion-ion and electron-ion pair distribution
functions, $g_{ij}(r)$ and $g_{ie}(r)$ respectively.  $g_{ij}(r)$ is simply related the
the pair correlation functions $h_{ij}(r)$
\begin{equation}
g_{ij}(r) = 1 + h_{ij}(r) 
\end{equation}
The all-electron, electron-ion pair distribution functions are given by
\begin{equation}
g_{ie}(r) = \frac{ n_{i,e}^{all}(r) }{ n_e^{all} }
\end{equation}
where
\begin{equation}
n_e^{all} =  \sum\limits_{\lambda=1}^{N} n_\lambda^0 Z_\lambda
\end{equation}
is the average electron density of all electrons in the plasma (including bound electrons), $Z_{\lambda}$ is the nuclear 
charge of species $\lambda$, and the spherically averaged electron density about a nucleus of species $i$ is
\begin{equation}
n_{i,e}^{all}(r) \equiv
 n_{i,e}^{PA}(r) + \sum\limits_{\lambda=1}^{N} n_\lambda^0 
\int d^3r^\prime g_{i\lambda}(|\br - \brp|) n_{\lambda, e}^{PA}(\brp)
\end{equation}
This last equation is a generalization of equation (\ref{dnie}) to all-electrons
and written in real space.

\section{Comparison with DFT-MD simulations}
\label{sec3}
All calculations with the new model have been carried out using the Dirac exchange functional \cite{dirac} and
the HNC closure relation in which $B_{II}=B_{IJ}=B_{JJ}=0$.  This latter approximation is not necessary 
since bridge function approximations exist \cite{faussurier1, rosenfeld, ocp_bridge, yukawa_bridge}, 
but is adequate for the present purposes.  The bridge function will become important for strong coupling cases.

In figure \ref{fig_ch_tf} we compare ion-ion pair distribution functions $g_{ij}(r)$ 
from the new model in the TF approximation (IS-TF) to OFMD simulations for CH$_{1.36}$, a material
related to the glow discharge polymer used in inertial confinement fusion experiments \cite{hammel}.
The OFMD simulations use 
the same exchange potential \cite{dirac} and the TF functional, and so are directly, comparable to the IS-TF calculations,  
and were carried out using 250 particles (106 Carbon + 144 Hydrogen) in a cubic cell with periodic boundary conditions.
The Hartree potential is calculated with FFT on a regular grid with 64$^3$ grid points and the time step is 0.01$\omega_p$, where
$\omega_p$ is the plasma frequency.
The agreement between IS-TF and OFMD is very good to excellent for all densities and temperatures, for all
three pair distribution functions (C-C, C-H and H-H).  The largest difference is seen for the highest density and lowest 
temperature, where the ion-ion coupling is the strongest.  The differences seen in this regime likely stem from the use of the HNC ion-ion
closure relation, which becomes inaccurate for strongly coupled cases.

In figure \ref{fig_ch_tf_gie} ion-electron pair distribution
functions $g_{ie}(r)$ corresponding to the lowest density calculations in figure \ref{fig_ch_tf} are shown.
Again, the agreement between the IS-TF model and the OFMD simulations is
excellent for all three temperatures.  The OFMD simulations use a pseudopotential which removes the Coulombic divergence
in the electron-ion interaction potential and replaces it with a well behaved, but artificial, pseudo-potential \cite{lambert3}.  For this reason 
we do not expect the OFMD electron densities to be accurate for $r< r_{c}$, where 
the actual potential has been replaced by the pseudopotential.  
The OFMD simulations return a finite value for $g_{ie}(r)$ at $r=0$, in contrast to the IS-TF results which exhibit the correct divergent
behavior ($\lim_{r\to 0} g_{ie}(r) \to r^{-3/2}$).  This highlights an advantage of the IS-TF method over OFMD, i.e.
it is an all-electron model so no pseudopotential is needed.

Next we turn to comparison between IS-QM and DFT-MD where both use the Kohn-Sham functional for the same
cases of a CH$_{1.36}$ mixture.  
Our QMD
simulations are performed using the 
Vienna ab initio Simulation Package (VASP)~\cite{vasp4}. 
Born-Oppenheimer MD within the NVT-ensemble with a Nos\'e-Hoover thermostat~\cite{nose1, nose2} is used throughout. 
We use a time step of 0.2 fs except for the very highest temperatures of our study, where we need to use a 0.1 fs time step 
in order to converge the internal energy and pressure to the desired accuracy. We use the generalized gradient approximation 
(GGA) of DFT with the  Perdew-Burke-Ernzerhof (PBE)~\cite{pbe1,pbe2} exchange-correlation functional. Projector-augmented 
wave (PAW)~\cite{paw1, paw2} pseudopotentials are used to account for the core electrons. We used the harder potentials 
for C and H in the VASP PAW library (with core radii of 1.1 and 0.8 a.u. respectively). The plane-wave cutoff is set to 1300 eV. 
The electronic density is constructed from single-particle wave functions by sampling only at the 
$(\frac{1}{4},\frac{1}{4},\frac{1}{4})$ of the Brillouin zone. The CH$_{1.36}$ mixture was simulated with 236 atoms (100 C and 136 H) which 
is close to the stochiometry of the plastic ablator used in inertial confinement capsules \cite{hamel-ch}. Pure carbon simulations  (figure \ref{fig_purec})
consisted of 64 atoms.

The comparison of ion-ion pair distribution functions is shown in figure \ref{fig_ch2}.
The agreement is very good for higher temperatures and higher densities, but degrades considerably at the lower
temperatures and densities.  To elucidate the origin of this disagreement we compare in figure \ref{fig_purec}
ion-ion pair distributions for pure carbon (i.e. not a mixture), over a similar range of conditions.
A similar trend in agreement with respect to density and temperature is seen.  This indicates
that it is not the extension of our model to mixtures that is the source of the disagreement, but rather that the
approximations involved in IS-QM are breaking down at the lowest temperatures and densities for carbon.  
Angular distribution functions for the pure carbon cases as calculated with QMD (not shown) indicate 
angular preference in nearest neighbor positions for the lower temperatures and densities in figure \ref{fig_purec}.
This implies that significant bonding between the carbon atoms remains at these temperatures and densities.
Such bonding is not captured in the IS-QM model which approximates the electron density as a superposition
of spherically symmetric pseudoatoms.
In contrast to this disagreement, in \cite{starrett13} 
good agreement with QMD simulations for aluminum was found at solid density (2.7g/cm$^3$) and down to 2eV (23.2kK).
Clearly the fact that aluminum forms a simple liquid favors the superposition approximation.  While for carbon and CH$_{1.36}$, strong bonding
and angular effects must be overcome by increasing the density or temperature before the model becomes accurate.
This is supported by the fact that there is good agreement with OFMD simulations for CH$_{1.36}$, for the same conditions,
where the use of the TF functional in the OFMD simulations precludes bonding effects (though not other angular effects).
The excellent agreement between IS-QM and QMD for CH$_{1.36}$ at 15g/cm$^3$ at 100kK
is expected to continue to higher densities and temperatures.  For higher temperatures than shown QMD quickly
becomes prohibitively expensive, whereas IS-QM remains tractable up to thousands of eV.

\section{Conclusions}
\label{sec4}
A model for the rapid calculation of the electronic and ionic structures of warm and hot dense
mixtures has been presented.  This is an extension of a previous model \cite{starrett13, starrett14}
for homo-nuclear plasmas.   Comparisons with DFT-MD simulations for CH$_{1.36}$ demonstrate excellent
agreement on ionic and electronic structure for Thomas-Fermi based calculations, while agreement
between Kohn-Sham based calculations is excellent for higher temperatures and densities but poor
for lower temperatures and densities.  A similar result is found by comparing
Kohn-Sham based calculations for a pure carbon plasma, under similar conditions.  Such disagreement 
had not been previously observed in comparisons for a pure aluminum plasma \cite{starrett13}.  This
is explained as a breakdown of the superposition approximation which underpins the model, as it
cannot describe the bonding observed in the Kohn-Sham DFT-MD simulations.

These initial comparisons indicate that the method is a promising technique for
calculating electronic and ionic structures where bonding is not significant.  In addition
the method is very rapid relative to the corresponding DFT-MD simulations.  Thomas-Fermi based calculations with the new model
take a few minutes on an single processor while Kohn-Sham based calculations can take a few hours\footnote{Our IS-QM calculations
use a shared memory, OpenMP parallel routine to solve for the continuum states \cite{starrett14}.}.  Moreover, the model is all 
electron, meaning that no pseudopotential is used, in contrast to DFT-MD simulations.
We note that the Kohn-Sham version of the 
model can access the high temperature regime (1000's of eV), something that is not computationally
tractable with Kohn-Sham DFT-MD.  Another advantage of the model is that highly asymmetric mixtures
(i.e. high mass ratios, charge ratios or extreme mixing fractions) present no additional difficulties, in contrast
with DFT-MD simulations.  Finally we note that the model could used as the basis for predicting 
X-ray scattering spectra \cite{souza14} and for calculating ionic transport properties
on the basis of effective potential theory \cite{baalrud13}.

\section*{Acknowledgments}
This work was performed under the auspices of the United States Department of Energy under contract 
DE-AC52-06NA25396 and LD-RD grant number 20130244ER.

\appendix

\section{Fourier transform definitions}
\label{ft}
Our convention for the Fourier transform of a spherically symmetric function $f(r)$ is
\begin{eqnarray}
f(k) & = & \frac{4 \pi }{k} \int\limits_0^\infty r f(r) \sin(kr) dr \\
\end{eqnarray}
and the inverse transform is
\begin{eqnarray}
f(r) & = & \frac{1 }{2 \pi^2 r} \int\limits_0^\infty k f(k) \sin(kr) dk.
\end{eqnarray}

\section{Numerical solution of the N-component OZ equations\label{noz_num}}
The $N$-component OZ equations (\ref{coz}) can be written in matrix form (in k-space)
\begin{eqnarray}
\ul{H} & = & \ul{C} +  \ul{C}\cdot \ul{D}\cdot \ul{H}
\end{eqnarray}
where $\ul{H}$ is the matrix of the pair correlation functions, $\ul{C}$ is the
direct correlation function matrix and $\ul{D}$ is a diagonal matrix of the particle
densities $n_\lambda^0$.  This can be rewritten as
\begin{eqnarray}
\ul{N} & = & \left( \ul{I} - \ul{C} \cdot \ul{D} \right)^{-1} \cdot \ul{C}\cdot \ul{D}\cdot \ul{C}
\label{n_matrix}
\end{eqnarray}
where $\ul{I}$ is the identity matrix and the nodal matrix is defined by
\begin{eqnarray}
\ul{N} & \equiv &  \ul{H} - \ul{C}
\end{eqnarray}
Equation (\ref{n_matrix}) allows us to proceed with the numerical solution in a straightforward
generalization of the 1-component procedure \cite{ng}.  Using the HNC closure relation and defining
$\ul{V}$ as the matrix of pair interaction potentials, the algorithm is as follows
\begin{enumerate}
\item
Inputs: \ul{V}, \ul{D} and $\alpha$ (a linear mixing parameter)
\item
Initial guess: $\ul{N}^0 = \ul{0}$
\item
Get $\ul{C}(k)$:
\begin{eqnarray}
\ul{H}(r) & = & \exp\left( -\beta \ul{V} + \ul{N} \right) \label{hm1} \\
\ul{C}(r) &  = &  \ul{H} - \ul{N} \nonumber\\
\ul{C}(k) &  = &  \Pi_f\left[\ul{C}(r)\right] \label{ft1}
\end{eqnarray}
where equations (\ref{hm1}) and  (\ref{ft1}) are element-wise operations and $\Pi_f$ represents the Fourier transform.
\item
Get new $\ul{N}$:
\begin{eqnarray}
\ul{N}(k) & = & \left( \ul{I} - \ul{C} \cdot \ul{D} \right)^{-1} \cdot \ul{C}\cdot \ul{D}\cdot \ul{C} \nonumber\\
\ul{N}(r)^{new} &  = &  \Pi_f^{-1} \left[\ul{N}(k)\right] \label{ft2}
\end{eqnarray}
(again the inverse Fourier transform $\Pi_f^{-1}$ is element-wise).  Get new trial $\ul{N}$ (linear mixing):
\begin{eqnarray}
\ul{N}^{i+1}(r) & = & (1-\alpha) \times \ul{N}^i + \alpha \times \ul{N}^{new}\nonumber
\end{eqnarray}
\item
Test convergence of new $\ul{N}^{i+1}$ against old $\ul{N}^i$.  If not converged go back to step 3, and
iterate until converged.
\end{enumerate}
The linear mixing parameter $\alpha$ is typically taken to be 0.1, but can be larger for weakly coupled cases, or
smaller for more strongly couple cases.  A smaller value gives more stable convergence for all cases, but solution takes longer.
However the overall cost of solving the OZ equations is still small, typically at most a few minutes.

\bibliographystyle{unsrt}
\bibliography{bibfile}

\end{document}